\documentclass[review]{elsarticle}
\journal{Journal of Statistical Planning and Inference}

\usepackage{natbib}
\usepackage{float}
\usepackage{graphicx}
\usepackage[dvipsnames]{xcolor}
\usepackage{footmisc}
\usepackage{amsmath}
\usepackage{amsthm}
\usepackage{amsfonts}
\usepackage{amssymb}
\usepackage{hyphenat}
\newtheorem{theorem}{Theorem}

\newtheorem{lemma}{Lemma}

\newtheorem{proposition}{Proposition}

\begin{document}

\begin{frontmatter}

\title{On Benjamini-Hochberg procedure applied to mid p-values}

\author[mymainaddress]{Xiongzhi Chen\corref{mycorrespondingauthor}}
\cortext[mycorrespondingauthor]{Corresponding author}
\ead{xiongzhi.chen@wsu.edu}

\author[mysecondaryaddress]{Sanat K. Sarkar}
\ead{sanat@temple.edu}

\address[mymainaddress]{Department of Mathematics and Statistics, Washington State University, Pullman, WA 99164, USA}
\address[mysecondaryaddress]{Department of
Statistical Science and Fox School of Business, Temple University,
Philadelphia, PA 19122, USA}

\begin{abstract}
Multiple testing with discrete p-values routinely arises in various scientific
endeavors. However, procedures, including the false discovery rate (FDR)
controlling Benjamini-Hochberg (BH) procedure, often used in such settings,
being developed originally for p-values with continuous distributions, are too
conservative, and so may not be as powerful as one would hope for. Therefore,
improving the BH procedure by suitably adapting it to discrete p-values
without losing its FDR control is currently an important path of research.
This paper studies the FDR control of the BH procedure when it is applied to
mid p-values and derive conditions under which it is conservative. Our
simulation study reveals that the BH procedure applied to mid p-values may be
conservative under much more general settings than characterized in this work,
and that an adaptive version of the BH procedure applied to mid p-values is as
powerful as an existing adaptive procedure based on randomized p-values.
\end{abstract}


\end{frontmatter}


\section{Introduction}

\label{secIntro}

Multiple testing based on discrete test statistics aiming at false discovery
rate (FDR) control has been widely conducted in many fields; see, e.g.,
\cite{Chen:2015discretefdr} and references therein. Knowing that many FDR
procedures, e.g., the Benjamini-Hochberg (BH) procedure in
\cite{Benjamini:1995} and Storey's procedure in \cite{Storey:2004}, tend to be
less powerful when applied to discrete p-values, three lines of research have
been attempted to address this issue. Among them, one is based on randomized
p-values as in the work of \cite{Habiger:2015}. Since randomized p-values are
uniformly distributed marginally, multiple testing based on such p-values are
essentially routed back to the continuous setting. However, results of
multiple testing based on randomized p-values may not be reproducible or
stable due to the use of randomized decision rules. On the other hand, mid
p-values \citep{Lancaster1961} are smaller than conventional p-values almost
surely, and a multiple testing procedure (MTP) may have larger power when
applied to mid p-values than conventional ones. However, there does not seem
to be a formal study on the BH procedure applied to mid p-values.

In this article, we focus on the FDR control of the BH procedure applied to
two-sided mid p-values of Binomial tests (BT's) and Fisher's exact tests
(FET's). Since mid p-values are not super-uniform, we derive simple conditions
under which the BH procedure is conservative in these settings. Compared to
multiple testing with p-values that are super-uniform, these conditions are
new and depict the critical role of the proportion of true null hypotheses for
FDR control when the cumulative distribution functions (CDF's) of p-values are
c\`{a}dl\`{a}g in general. In particular, they explicitly show the
interactions between the supremum norms of the probability density functions
(PDF's) of p-values, the proportion of true null hypotheses, the nominal
FDR\ level and the number of hypotheses to test in order to ensure the
conservativeness of the BH procedure applied to two-sided mid p-values. Our
simulation study provides strong numerical evidence on the conservativeness
and improved power of the BH procedure applied to mid p-values.

The rest of the article is organized as follows. \autoref{SecPrelim}
introduces some notations, three definitions of two-sided p-value and the
setting for multiple testing based on p-values. \autoref{SecTwoTypes}
discusses FDR bounds for step-up procedures based on p-values with
c\`{a}dl\`{a}g CDF's and those for the BH procedure applied to two-sided mid
p-values. \autoref{SecSimuStudy} presents a simulation study on the BH
procedure and its adaptive version for mid p-values and conventional p-values. \autoref{seAppAll}
provides an application of the BH based on two-sided mid p-values to an HIV study.
\autoref{SecDiscussion} ends the article with a discussion.

\section{Preliminaries}

\label{SecPrelim}

\subsection{Notations and conventions}

Any CDF is assumed to be right-continuous with left-limits, i.e.,
c\`{a}dl\`{a}g, and the set of CDF's is denoted by $\mathcal{D}$. For any
$F\in\mathcal{D}$, denote its support by $S_{F}$. For a real-valued function
$g$ with domain $D$, $\left\Vert g\right\Vert _{\infty}=\sup_{x\in
D}\left\vert g\left(  x\right)  \right\vert $. \textquotedblleft if and only
if\textquotedblright\ will be abbreviated as \textquotedblleft
iff\textquotedblright. $\left[  x\right]  $ denotes the integer part of
$x\in\mathbb{R}$.

\subsection{Three definitions of a two-sided p-value}

\label{SecDefinition}

For a random variable $X$, let $F$ be its CDF with support $\mathcal{S}$ and
$f$ be its PDF defined as the Radon-Nikodym derivative $\frac{\mathsf{d}%
F}{\mathsf{d}\upsilon}$ with $\upsilon$ being the Lebesgue measure or the
counting measure on $\mathcal{S}$. For an observation $x_{0}$ from $X$, set%
\[
l\left(  x_{0}\right)  =\int_{\left\{  x\in\mathcal{S}:f\left(  x\right)
<f\left(  x_{0}\right)  \right\}  }\mathsf{d}F\left(  x\right)  \text{ \ \ and
\ }e\left(  x_{0}\right)  =\int_{\left\{  x\in\mathcal{S}:f\left(  x\right)
=f\left(  x_{0}\right)  \right\}  }\mathsf{d}F\left(  x\right)  .
\]
Based on \cite{Agresti:2002}, a two-sided conventional p-value for $x_{0}$ is
defined as $p\left(  x_{0}\right)  =l\left(  x_{0}\right)  +e\left(
x_{0}\right)  $. It is well known that $\Pr\left(  p\left(  X\right)  \leq
t\right)  \leq t$ for all $t\in\left[  0,1\right]  $ and $\Pr\left(  p\left(
X\right)  \leq p\left(  x\right)  \right)  =p\left(  x\right)  $ for all
$x\in\mathcal{S}$. Using Theorem 2 of \cite{Dickhaus:2012}, the two-sided
randomized p-value is defined as $\rho\left(  x_{0},u\right)  =l\left(
x_{0}\right)  +\left(  1-u\right)  e\left(  x_{0}\right)  $, where $u$ is a
realization of $U\sim\mathsf{Uniform}\left(  0,1\right)  $, i.e., the uniform
random variable on $\left[  0,1\right]  $ and $U$ is independent of $X$. Note
that $\rho\left(  X,U\right)  \sim\mathsf{Uniform}\left(  0,1\right)  $
marginally. Following \cite{Hwang:2001}, the two-sided mid p-value is defined
as $\varpi\left(  x_{0}\right)  =l\left(  x_{0}\right)  +2^{-1}e\left(
x_{0}\right)  $. Note that $\varpi$ has some optimality properties justified
by \cite{Hwang:2001}. Throughout this article, $P$ is the generic symbol for
p-value, which can be $p$, $\rho$ or $\varpi$.

A random variable $Y$ with range in $\left[  0,1\right]  $ is called
\textquotedblleft super-uniform\textquotedblright\ if $\Pr\left(  Y\leq
t\right)  \leq t$\ for all $t\in\left[  0,1\right]  $, and it is called
\textquotedblleft sub-uniform\textquotedblright\ if $\Pr\left(  Y\leq
t\right)  > t$\ for all $t$ in the support of its distribution.

\begin{lemma}
\label{Sunif}For any $x\in\mathcal{S}$,%
\begin{equation}
\Pr\left(  \varpi\left(  X\right)  \leq\varpi\left(  x\right)  \right)
=p\left(  x\right)  =\varpi\left(  x\right)  +2^{-1}e\left(  x\right)  .
\label{eq15}%
\end{equation}
Further, $\mathbb{E}\left[  \left.  \rho\left(  X,U\right)  \right\vert
X\right]  =\varpi\left(  X\right)  $. Finally, assume $\left\{  u_{j}\right\}
_{j=1}^{n}$ are i.i.d. $\mathsf{Uniform}\left(  0,1\right)  $ and independent
of $X$ and let $\rho\left(  X,u_{j}\right)  =l\left(  X\right)  +\left(
1-u_{j}\right)  e\left(  X\right)  $. Then, conditional on $X$,%
\begin{equation}
\lim_{n\rightarrow\infty}\frac{1}{n}\sum_{j=1}^{n}\rho\left(  X,u_{j}\right)
=\varpi\left(  X\right)  \ \text{almost surely.} \label{eqd1}%
\end{equation}

\end{lemma}

\begin{proof}
Identity (\ref{eq15}) holds due to%
\[
\Pr\left(  \varpi\left(  X\right)  \leq\varpi\left(  x\right)  \right)
=\int_{\left\{  x^{\prime}\in\mathcal{S}:f\left(  x^{\prime}\right)  \leq
f\left(  x\right)  \right\}  }\mathsf{d}F\left(  x\right)
\]
and the definitions of $p$, $\varpi$ and $e$. The validity of $\mathbb{E}%
\left[  \left.  \rho\left(  X,U\right)  \right\vert X\right]  =\varpi\left(
X\right)  $ follows from%
\begin{align*}
\mathbb{E}\left[  \left.  \rho\left(  X,U\right)  \right\vert X\right]   &
=\mathbb{E}\left[  \left.  l\left(  X\right)  +\left(  1-U\right)  e\left(
X\right)  \right\vert X\right] \\
&  =l\left(  X\right)  +2^{-1}e\left(  X\right)  =\varpi\left(  X\right)  ,
\end{align*}
where we have used the independence between $U$ and $X$ to obtain the second
equality. Finally, (\ref{eqd1}) holds by the mutual independence between
$\left\{  u_{j}\right\}  _{j=1}^{n}$ and $X$ and the strong law of large
numbers. This completes the proof.
\end{proof}

\autoref{Sunif} implies that $\varpi$ is sub-uniform. However, for a two-sided mid p-value whose CDF is not a Dirac mass, the set on which it is strictly super-uniform, i.e., the set $S_{\textrm{su}}=\left\{t^{\prime} \in \left[0,1\right]: \Pr\left(\varpi \le t^{\prime}\right) < t^{\prime}\right\}$, is non-empty and is the union of disjoint sub-intervals of $\left[0,1\right]$. Another implication of
\autoref{Sunif} is that, averaging a large number of realizations of a random
p-value $\rho$ in order to reduce its extra uncertainty induced by $U$
essentially makes $\rho$ into a mid p-value $\varpi$. In other words, the
stability and reproducibility issues of multiple testing based on randomized
p-values is incompatible with its key motivation.

\subsection{Multiple testing based on p-values}

In a typical multiple testing setting, there are $m$ null hypothesis $\left\{
H_{i}\right\}  _{i=1}^{m}$, among which $m_{0}$ are true nulls and the rest
$m_{1}$ false nulls. Further, a p-value $P_{i}$ is associated with $H_{i}$ for
each $i$, and an MTP is usually applied to $\left\{  P_{i}\right\}  _{i=1}%
^{m}$. Let $I_{0}$ be the index set of true nulls and $I_{1}$ be the
complement of $I_{0}$. Then the proportion of true nulls $\pi_{0}$ is defined
as $m_{0}/m$ and that of false nulls $\pi_{1}$ as $1-\pi_{0}$.

Let $\left\{  P_{\left(  i\right)  }\right\}  _{i=1}^{m}$ be the ordered
version of $\left\{  P_{i}\right\}  _{i=1}^{m}$ such that $P_{\left(
1\right)  }\leq P_{\left(  2\right)  }\leq\cdots\leq P_{\left(  m\right)  }$,
and $H_{\left(  i\right)  }$ the null hypothesis associated with $P_{\left(
i\right)  }$ for each $i$. A step-up MTP with critical constants $\left\{
\tau_{i}\right\}  _{i=1}^{m}$ such that $0<\tau_{i}\leq\tau_{i+1}\leq1$ for
$1\leq i\leq m-1$ rejects $H_{\left(  j\right)  }$ when $P_{\left(  j\right)
}\leq\tau_{\eta}$ if%
\[
\eta=\max\left\{  1\leq i\leq m:P_{\left(  i\right)  }\leq\tau_{i}\right\}
\]
exists, and rejects no null hypothesis otherwise. For an MTP, let $V$ be the
number of false discoveries, i.e., the number of true nulls that are rejected,
and $R$ the number of rejected nulls. Then the FDR\ of the MTP is defined as
$\mathbb{E}\left(  \frac{V}{\max\left\{  R,1\right\}  }\right)  $. The BH
procedure is the step-up MTP with $\tau_{i}=i\alpha/m$ for $1\leq i\leq m$ and
is designed to control its FDR at level $\alpha\in\left(  0,1\right)  $.

\section{Non-asymptotic FDR bounds under independence}

\label{SecTwoTypes}

In this section, we will derive FDR upper\ bounds for a step-up procedure when
p-values are independent and have c\`{a}dl\`{a}g CDF's, and then provide
conditions on the conservativeness of the BH\ procedure when it is applied to
mid p-values.

Let $\alpha\in\left(  0,1\right)  $ be the nominal FDR level and consider a
step-up procedure with critical constants $\left\{  \tau_{i}\right\}
_{i=1}^{m}$. Let $\hat{\alpha}$ be the FDR\ of the procedure. For each $i\in
I_{0}$ and $r\in\left\{  1,\ldots,m\right\}  $, let $C_{r}^{\left(  -i\right)
}$ be the event that if $H_{i}$, $i\in I_{0}$ is rejected, then $r-1$
hypotheses among $\left\{  H_{j}:j\neq i\right\}  $ are rejected. This yields
the following representation
\begin{equation}
\hat{\alpha}=\sum_{i\in I_{0}}\sum_{r=1}^{m}\frac{1}{r}\Pr\left(  p_{i}%
\leq\tau_{r},C_{r}^{\left(  -i\right)  }\right)  \label{eq7}%
\end{equation}
as in \cite{Benjamini:2001}; see also \cite{Sarkar:2008b}, where an explicit
expression is given for $C_{r}^{(-i)}$ in terms of the step-up procedure using
$\{H_{j}:j\neq i\}$ and the critical constants $\{\tau_{i}\}_{i=2}^{m}$.

For each $i$, let $F_{i}$ be the CDF of $P_{i}$ obtained by assuming $H_{i}$
is a true null. We call $F_{i}$ the null distribution of $P_{i}$, and denote
by $S_{i}$ the support of $F_{i}$.

\begin{lemma}
\label{lmGeneral}If $\left\{  P_{i}\right\}  _{i=1}^{m}$ are independent, then%
\begin{equation}
\hat{\alpha}=\sum_{i\in I_{0}}\sum_{r=1}^{m}\frac{1}{r}F_{i}\left(  \tau
_{r}\right)  \Pr\left(  C_{r}^{\left(  -i\right)  }\right)  . \label{eq9}%
\end{equation}
If in addition%
\begin{equation}
\max_{1\leq r\leq m}\max_{i\in I_{0}}r^{-1}F_{i}\left(  \tau_{r}\right)
\leq\frac{\alpha}{m_{0}}, \label{eq8}%
\end{equation}
then $\hat{\alpha}\leq\alpha$.
\end{lemma}

Expression (\ref{eq9}) follows from (\ref{eq7}) and the independence
assumption, and (\ref{eq8}) follows from the fact that
\[
\sum_{r=1}^{m}\Pr\left(  C_{r}^{\left(  -i\right)  }\right)  =1,\forall i\in
I_{0}.
\]
When each $P_{i},i\in I_{0}$ is super-uniform and $\tau_{i}=\frac{i\alpha}{m}$
for each $i$, the inequality (\ref{eq8}) becomes%
\[
\max_{1\leq r\leq m}\max_{i\in I_{0}}\frac{1}{r}F_{i}\left(  \tau_{r}\right)
\leq\frac{\alpha}{m}\leq\frac{\alpha}{m_{0}},
\]
which recovers the fact that the BH procedure is conservative.

To avoid unnecessary complications in dealing with maxima and suprema, in the
rest of the article we will only consider $F$ whose $S_{F}\ $is finite. For
any fixed $t\in(0,1]$, define%
\[
\xi\left(  t\right)  =\operatorname*{argmin}\left\{  t-P\left(  s\right)
:s\in S,P\left(  s\right)  \leq t\right\}  ,
\]
i.e., $\xi\left(  t\right)  $ is the set of observations of $X$ whose p-values
are the closest to $t$. Note that $\xi\left(  t\right)  =0$ and $e\left(
\xi\left(  t\right)  \right)  =0$ are set when $\left\{  s\in S:P\left(
s\right)  \leq t\right\}  $ is empty. Recall $S_{i}$ as the support of $P_{i}$
and let $f_{i}$ be the PDF of $P_{i}$. For any $t\in\left(  0,1\right)  $ and
each $i$, let%
\[
l_{i}\left(  x^{\prime}\right)  =\int_{\left\{  x\in S_{i}:f_{i}\left(
x\right)  <f_{i}\left(  x^{\prime}\right)  \right\}  }\mathsf{d}F_{i}\left(
x\right)  \text{ \ \ and \ }e_{i}\left(  x^{\prime}\right)  =\int_{\left\{
x\in S_{i}:f_{i}\left(  x\right)  =f_{i}\left(  x^{\prime}\right)  \right\}
}\mathsf{d}F_{i}\left(  x\right)
\]
for $x^{\prime}\in S_{i}$ and%
\[
x_{i}\left(  t\right)  =\operatorname*{argmin}\left\{  t-P_{i}\left(
s\right)  :s\in S_{i},P_{i}\left(  s\right)  \leq t\right\}  .
\]

\begin{lemma}
\label{LmAntiCons}Assume $\left\{  \varpi_{i}\right\}  _{i=1}^{m}$ are
independent. Then the FDR\ $\hat{\alpha}_{\mathsf{BH}}$ of the BH procedure
satisfies%
\[
\hat{\alpha}_{\mathsf{BH}}=\sum_{i\in I_{0}}\sum_{r=1}^{m}\frac{1}{r}\left(
\tau_{r}+2^{-1}e_{i}\left(  x_{i}\left(  \tau_{r}\right)  \right)  \right)
\Pr\left(  C_{r}^{\left(  -i\right)  }\right)
\]
when it is applied to $\left\{  \varpi_{i}\right\}  _{i=1}^{m}$.
\end{lemma}

The proof of \autoref{LmAntiCons} follows immediately from (\ref{eq15}),
(\ref{eq7}) and (\ref{eq9}) and is omitted. \autoref{LmAntiCons} implies that
the BH procedure is not conservative when $\pi_{0}=1$ when it is applied to
two-sided mid p-values, and it suggests that the BH critical constants are
tight for weak familywise error rate (FWER) control in the stochastic order of
p-values with respect to the uniform random variable. In the rest of this
section, we consider FDR\ bounds for multiple testing based on two-sided mid
p-values $\left\{  \varpi_{i}\right\}  _{i=1}^{m}$ of BT's and FET's when
$\pi_{0}<1$.

\subsection{Bounds associated with mid p-values of Binomial tests}

The Binomial test (BT) is used to test if two independent Poisson distributed
random variables, $X_{i}\sim\mathsf{Poisson}\left(  \lambda_{i}\right)
,i=1,2$, have the same mean parameters $\lambda_{i}$. Let $\mathsf{Binomial}%
\left(  \theta_{\ast},c_{\ast}\right)  $ denote a Binomial distribution with
probability of success $\theta_{\ast}$ and total number of trials $c_{\ast}$.
Suppose a count $c_{i}$ is observed from $X_{i}$, then the BT statistic
$T_{\theta_{i}}\sim\mathsf{Binomial}\left(  \theta_{i},c\right)  $ with
$\theta_{i}=\lambda_{i}\left(  \lambda_{1}+\lambda_{2}\right)  ^{-1}$ and
$c=c_{1}+c_{2}$. Under the null $H_{0}:\lambda_{1}=\lambda_{2}$, we have
$\theta=0.5$ for $i=1,2$. Given $c_{1}$ or $c_{2}$, the two-sided p-value
associated with $T_{\theta}$ is computed using the CDF of $T_{0.5}$. Note that
the PDF of $\mathsf{Binomial}\left(  0.5,n\right)  $ is simply $f\left(
x;n\right)  =\binom{n}{x}2^{-n}$ for $x=0,1,\ldots,n$.

\begin{lemma}
\label{LmBinomialCDF}Let $n$ and $n^{\prime}$ be two positive integers such
that $n^{\prime}>n$ and $x\in\left\{  0,\ldots,n\right\}  $. Then
$\frac{f\left(  x;n\right)  }{f\left(  x;n^{\prime}\right)  }<1$
if\ $x>2^{-1}n^{\prime}$, and $\frac{f\left(  x;n\right)  }{f\left(
x;n^{\prime}\right)  }>1$ if\ $x<2^{-1}\left(  n+1\right)  $. Further,
$\operatorname*{argmax}_{0\leq x\leq n}f\left(  x;n\right)  =\left\{
\frac{n-1}{2},\frac{n+1}{2}\right\}  $ when\ $n$ is odd, and
$\operatorname*{argmax}_{0\leq x\leq n}f\left(  x;n\right)  =\left[
\frac{n+1}{2}\right]  $ when $n$ is even. Therefore, $\frac{\left\Vert
f\left(  \cdot;n\right)  \right\Vert _{\infty}}{\left\Vert f\left(
\cdot;n+1\right)  \right\Vert _{\infty}}=\frac{n+2}{n+1}$ for $n$ even\ and
$\frac{\left\Vert f\left(  \cdot;n\right)  \right\Vert _{\infty}}{\left\Vert
f\left(  \cdot;n+1\right)  \right\Vert _{\infty}}=1$ for $n$\ odd.
\end{lemma}

\begin{proof}
Since%
\[
\frac{f\left(  x;n\right)  }{f\left(  x;n^{\prime}\right)  }=2^{n^{\prime}%
-n}\left(  1-\frac{x}{n^{\prime}}\right)  \left(  1-\frac{x}{n^{\prime}%
-1}\right)  \cdots\left(  1-\frac{x}{n+1}\right)  ,
\]
we see%
\[
2^{n^{\prime}-n}\left(  1-\frac{x}{n+1}\right)  ^{n^{\prime}-n}\leq
\frac{f\left(  x;n\right)  }{f\left(  x;n^{\prime}\right)  }\leq2^{n^{\prime
}-n}\left(  1-\frac{x}{n^{\prime}}\right)  ^{n^{\prime}-n}.
\]
So, $\frac{f\left(  x;n\right)  }{f\left(  x;n^{\prime}\right)  }<1$
if\ $x>2^{-1}n^{\prime}$, and $\frac{f\left(  x;n\right)  }{f\left(
x;n^{\prime}\right)  }>1$ if\ $x<2^{-1}\left(  n+1\right)  $, i.e., the first
claim holds. The second claim holds since%
\[
\frac{f\left(  x+1;n\right)  }{f\left(  x;n\right)  }=\frac{n-x}{x+1}%
\]
for $x\in\left\{  0,\ldots,n-1\right\}  $ and $\frac{f\left(  x+1;n\right)
}{f\left(  x;n\right)  }<1$ iff $x<\frac{n-1}{2}$, with equality iff
$x=\frac{n-1}{2}$. Finally, we show the third claim. Let $k$ be a non-negative
integer. When $n=2k$ for $k\geq1$,%
\[
\frac{\left\Vert f\left(  \cdot;n\right)  \right\Vert _{\infty}}{\left\Vert
f\left(  \cdot;n+1\right)  \right\Vert _{\infty}}=2\times\frac{\left(
2k\right)  !\left(  k+1\right)  !k!}{k!k!\left(  2k+1\right)  !}=\frac
{2k+2}{2k+1}.
\]
On the other hand, when $n=2k+1$ for $k\geq0$,%
\[
\frac{\left\Vert f\left(  \cdot;n\right)  \right\Vert _{\infty}}{\left\Vert
f\left(  \cdot;n+1\right)  \right\Vert _{\infty}}=2\times\frac{\left(
2k+1\right)  !\left(  k+1\right)  !\left(  k+1\right)  !}{k!\left(
k+1\right)  !\left(  2k+2\right)  !}=1\text{.}%
\]
This completes the proof.
\end{proof}

\autoref{LmBinomialCDF} implies that $f\left(  x;n\right)  $ dominates
$f\left(  x;n^{\prime}\right)  $ for $n^{\prime}>n$ and $x\leq2^{-1}\left(
n+1\right)  $ and that the maximum, $\left\Vert f\left(  \cdot;n\right)
\right\Vert _{\infty}$, of the PDF of $\mathsf{Binomial}\left(  0.5,n\right)
$ is non-increasing in $n$.

Now we consider applying the BH procedure to two-sided mid p-values of BT's
for multiple testing of equality of Poisson means. Assume there are $2m$ mutually
independent Poisson random variables, $\mathsf{Poisson}\left(  \lambda
_{si}\right)  $ for $s=1,2$ and $i=1,\ldots,m$, such that $\mathsf{Poisson}%
\left(  \lambda_{1i}\right)  $ and $\mathsf{Poisson}\left(  \lambda
_{2i}\right)  $ form a pair for each $i$. For each $i=1,\ldots,m$, a BT is
conducted to assess the null $H_{i}:\lambda_{1i}=\lambda_{2i}$ versus the
alternative $H_{i}^{\ast}:\lambda_{1i}\neq\lambda_{2i}$, and a two-sided mid
p-value $\varpi_{i}$ is obtained. Then the BH procedure is applied to
$\left\{  \varpi_{i}\right\}  _{i=1}^{m}$ to determine which null hypotheses
are true. In this setting, $\pi_{0}$ is the proportion among the $m$ pairs of
Poisson random variables that have equal means. For each $i$, denote the
distribution of the corresponding BT by $\mathsf{Binomial}\left(  \theta
_{i},n_{i}\right)  $, and write $f\left(  \cdot;n_{i}\right)  $ as
$f_{i}\left(  \cdot\right)  $.

\begin{proposition}
\label{PropMTPBt}Let $n_{\ast}=\min_{1\leq i\leq m}n_{i}$ and $i_{0}%
\in\operatorname*{argmin}_{1\leq i\leq m}n_{i}$. If $n_{\ast}>0$, $\left\{
\varpi_{i}\right\}  _{i=1}^{m}$ are independent, $\pi_{0}<1$ and
\begin{equation}
f_{i_{0}}\left(  x_{i_{0}}\left(  \alpha\right)  \right)  \leq\frac{\left(
1-\pi_{0}\right)  \alpha}{m_{0}}, \label{eq20}%
\end{equation}
then the BH procedure is conservative.
\end{proposition}

\begin{proof}
When $\tau_{i}=\frac{i\alpha}{m}$ for $1\leq i\leq m$ and $\alpha<1$, we see
that, for each $1\leq i\leq m$, $\max_{1\leq r\leq m}x_{i}\left(  \tau
_{r}\right)  $ is strictly less than the mode(s) of $f_{i}$ and is equal to
$x_{i}\left(  \alpha\right)  $ by symmetry of $f_{i}$ with respect to
$2^{-1}n_{i}$. So, $x_{i_{0}}\left(  \alpha\right)  $ is strictly smaller than
the mode(s) of $f_{i_{0}}$. However, \autoref{LmBinomialCDF} implies $f\left(
x;n_{\ast}\right)  >f\left(  x;n^{\prime}\right)  $ if\ $x<2^{-1}\left(
n_{\ast}+1\right)  $ for all $n^{\prime}>n_{\ast}$. Therefore, from
\autoref{LmAntiCons} we obtain%
\begin{align}
\hat{\alpha}  &  \leq\pi_{0}\alpha+\sum_{i\in I_{0}}\sum_{r=1}^{m}\frac{1}%
{r}f_{i_{0}}\left(  x_{i_{0}}\left(  \alpha\right)  \right)  \Pr\left(
C_{r}^{\left(  -i\right)  }\right) \nonumber\\
&  \leq\pi_{0}\alpha+m_{0}f_{i_{0}}\left(  x_{i_{0}}\left(  \alpha\right)
\right)  \label{eq67}%
\end{align}
since $\sum_{r=1}^{m}\Pr\left(  C_{r}^{\left(  -i\right)  }\right)  =1$ for
each $i\in I_{0}$. It is easy to verify that (\ref{eq67}) is bounded by
$\alpha$ when (\ref{eq20}) holds. This completes the proof.
\end{proof}

\autoref{PropMTPBt} implies that, when $m_{0}$ is known and less than $m$, it
suffices to check $f_{i_{0}}\left(  x_{i_{0}}\left(  \alpha\right)  \right)  $
corresponding to the test that has the smallest positive count, in order to
ensure the conservativeness of the BH procedure when it is applied to
$\left\{  \varpi_{i}\right\}  _{i=1}^{m}$. It also reveals that, compared to
multiple testing with super-uniform p-values, $\pi_{0}<1$ is critical for
FDR\ control when not all p-values are super-uniform. Note that condition
(\ref{eq20}) is easily satisfied when $m_0$ and $\pi_0$ are small and $n_{\ast}$ is relatively large.
For example, when $\alpha=0.05$, $\pi_0=0.2$ and $m_0 =2$, the upper bound in \eqref{eq20} becomes $0.02$,
and $n_{\ast}=120$, $122$ or $124$ validates \eqref{eq20} (whose corresponding left side quantity is $0.01896$, $0.01922$
or $0.01948$, respectively). However, we admit that condition \eqref{eq20} is restrictive.

\subsection{Bounds associated with mid p-values of Fisher's exact tests}

Fisher's exact test (FET) has been widely used in assessing if a discrete
conditional distribution is identical to its unconditional version, where the
observations are modelled by Binomial distributions. Suppose for each $i=1,2$
a count $c_{i}$ is observed from $X_{i}\sim\mathsf{Binomial}\left(
q_{i},N_{i}\right)  $. Then the marginal $\mathbf{N}=\left(  N_{1}%
,N_{2},M\right)  $ with $M=c_{1}+c_{2}$ as the total count is obtained, and
the test statistic $T_{\theta}$ of the FET follows a hypergeometric
distribution $\mathsf{HGeom}\left(  \theta,\mathbf{N}\right)  $ with PDF
\[
f\left(  x;\theta,\mathbf{N}\right)  =\left.  \binom{N_{1}}{x}\binom{N_{2}%
}{M-x}\theta^{x}\right/  \sum_{u=x_{\ast}}^{x^{\ast}}\binom{N_{1}}{u}%
\binom{N_{2}}{M-u}\theta^{u}%
\]
for $x_{\ast}\leq x\,\leq x^{\ast}$, $x_{\ast}=\max\left\{  0,M-N_{2}\right\}
,x^{\ast}=\min\left\{  N_{1},M\right\}  \ $and $\theta\in\left(
0,\infty\right)  $. We will write $f\left(  \cdot;\theta,\mathbf{N}\right)  $
as $f\left(  \cdot;\mathbf{N}\right)  $ when $\theta=1$. Under the null
hypothesis $H_{0}:q_{1}=q_{2}$, if $q_{1},q_{2}\in\left(  0,1\right)  $ then
$\theta=1$ holds. The two-sided p-value associated with $T_{\theta}$\ for the
observation $c_{1}$ or $c_{2}$ is defined using the CDF\ of $T_{1}$.

When $N_{1}=N_{2}$, the distribution of $T_{1}$ only depends on $M$, and
$f\left(  x;\theta,\mathbf{N}\right)  $ reduces to%
\[
f\left(  x;\mathbf{N}\right)  =\left.  \binom{N}{x}\binom{N}{M-x}\right/
\binom{2N}{M}%
\]
and is written as $f\left(  x;N,M\right)  $.

\begin{lemma}
\label{lmOrderHyper}Assume $\mathbf{N}=\left(  N,N,M\right)  $. Then
$\frac{f\left(  x;\mathbf{N}\right)  }{f\left(  x+1;\mathbf{N}\right)  }<1$
iff$\ x<\frac{M-1}{2},$ with equality iff $x=\frac{M-1}{2}$. So,
$\operatorname*{argmax}f\left(  x;\mathbf{N}\right)  =\left\{  \frac{M-1}%
{2},\frac{M+1}{2}\right\}  $ when\ $M$ is odd, and $\operatorname*{argmax}%
f\left(  x;\mathbf{N}\right)  =\left[  \frac{M+1}{2}\right]  $ when $M$ is
even. Let $\kappa_{N}\left(  M\right)  =\frac{\left\Vert f\left(
\cdot;N,M\right)  \right\Vert _{\infty}}{\left\Vert f\left(  \cdot
;N,M+1\right)  \right\Vert _{\infty}}$. Then $\kappa_{N}\left(  M\right)
=\frac{M+2}{M+1}>1$ if $M$ is even but $\kappa_{N}\left(  M\right)
=\frac{2N-M}{2N-M+1}<1$ when $M$ is odd. Further, $\frac{f\left(
x;N,M\right)  }{f\left(  x;N,M+1\right)  }>1\ $iff\ $x<\frac{\left(
M+1\right)  N}{2N+1},$ with equality iff $x=\frac{\left(  M+1\right)  N}%
{2N+1}$.
\end{lemma}

\begin{proof}
Recall $\mathbf{N}=\left(  N,N,M\right)  $. Then%
\[
\frac{f\left(  x;N,M\right)  }{f\left(  x+1;N,M\right)  }=\frac{\left(
x+1\right)  \left(  N-M+x+1\right)  }{\left(  M-x\right)  \left(  N-x\right)
},
\]
and $\frac{f\left(  x;N,M\right)  }{f\left(  x+1;N,M\right)  }<1$ iff
$x<\frac{M-1}{2}$, with equality iff $x=\frac{M-1}{2}$. This justifies the
first claim. We move to the second claim. Let $k$ be a non-negative integer.
Then, when $M=2k$ with $k\geq1$,%
\[
\kappa_{N}\left(  M\right)  =\frac{\binom{N}{k}}{\binom{2N}{2k}}\frac
{\binom{2N}{2k+1}}{\binom{N}{k+1}}=\frac{2k+2}{2k+1}>1
\]
and when $N=2k+1$ with $k\geq0$,%
\[
\kappa_{N}\left(  M\right)  =\frac{\binom{N}{k}}{\binom{N}{k+1}}\frac
{\binom{2N}{2k+2}}{\binom{2N}{2k+1}}=\frac{2N-2k-1}{2N-2k}<1.
\]
This justifies the second claim. Now we show the third claim. Note that
$N-M+x\geq0$ when $x_{\ast}\leq x\leq x^{\ast}$ by the definition of $f\left(
\cdot;N,M\right)  $. From%
\[
\frac{f\left(  x;N,M\right)  }{f\left(  x;N,M+1\right)  }=\frac{\left(
M+1-x\right)  \left(  2N-M\right)  }{\left(  M+1\right)  \left(  N-M+x\right)
},
\]
we see that $\frac{f\left(  x;N,M\right)  }{f\left(  x;N,M+1\right)  }>1$ iff
$x<\frac{\left(  M+1\right)  N}{2N+1}$, with equality iff $x=\frac{\left(
M+1\right)  N}{2N+1}$. This completes the proof.
\end{proof}

\autoref{lmOrderHyper} implies that the ratio $\kappa_{N}\left(  M\right)  $
of the supremum norms for the PDFs of $\mathsf{HGeom}\left(  1,\mathbf{N}%
\right)  $ with $N$ fixed zigzags around $1$ as $M$ changes from being odd to
even, and that $f\left(  x;N,M\right)  $ dominates $f\left(  x;N,M^{\prime
}\right)  $ when $x\leq\frac{\left(  M+1\right)  N}{2N+1}$ and $2N\geq
M^{\prime}>M$.

Now let us consider applying the BH procedure to two-sided mid p-values of
FET's for multiple testing of equality of probabilities of success of Binomial
random variables when their total number of trials are the same. Suppose there
are $2m$ mutually independent Binomial random variables, $\mathsf{Binomial}\left(  q_{si},N\right)  $ for
$s=1,2$ and $i=1,\ldots,m$, such that $\mathsf{Binomial}\left(  q_{1i}%
,,N\right)  $ and $\mathsf{Binomial}\left(  q_{2i},,N\right)  $ form a pair
for each $i$. For each $i=1,\ldots,m$, FET is conducted to assess the null
$H_{i}:q_{1i}=q_{2i}$ versus the alternative $H_{i}^{\ast}:q_{1i}\neq q_{2i}$,
and a two-sided mid p-value $\varpi_{i}$ is obtained. Then the
BH procedure is applied to $\left\{  \varpi_{i}\right\}  _{i=1}^{m}$ to
determine which null hypotheses are true.
In this setting,
$\pi_{0}$ is the proportion among the $m$ pairs of
Binomial random variables that have equal probabilities of success.
For each $i$, denote the
distribution of the corresponding FET by $\mathsf{HGeom}\left(  \theta
_{i},\mathbf{N}_{i}\right)  $ with $\mathbf{N}_{i}=\left(  N,N,M_{i}\right)  $
and write $f\left(  \cdot;\mathbf{N}_{i}\right)  $ as $f_{i}\left(
\cdot\right)  $.

\begin{proposition}
\label{PropMTPFET}Assume $\mathbf{N}_{i}=\left(  N,N,M_{i}\right)  $ for all
$i$. Let $M_{\ast}=\min_{1\leq i\leq m}M_{i}$ and $i_{0}\in
\operatorname*{argmin}_{1\leq i\leq m}M_{i}$. If $M_{\ast}>1$, $\left\{
\varpi_{i}\right\}  _{i=1}^{m}$ are independent, $\pi_{0}<1$ and%
\begin{equation}
f_{i_{0}}\left(  x_{i_{0}}\left(  \alpha\right)  \right)  \leq\frac{\left(
1-\pi_{0}\right)  \alpha}{m_{0}}, \label{eq30}%
\end{equation}
then the BH procedure is conservative.
\end{proposition}

The proof of \autoref{PropMTPFET} is very similar to that of
\autoref{PropMTPBt} and omitted. \autoref{PropMTPFET} implies that, when
$m_{0}$ is known and less than $m$, it suffices to check $f_{i_{0}}\left(
x_{i_{0}}\left(  \alpha\right)  \right)  $ corresponding to the test that has
the smallest positive total count, in order to ensure the conservativeness of
the BH procedure applied to $\left\{  \varpi_{i}\right\}  _{i=1}^{m}$. Similar
to the case of two-sided mid p-values of the BT's, condition (\ref{eq30}) is
easily satisfied when $m_0$ and $\pi_0$ are small and $n_{\ast}$ is relatively large.
For example, when $\alpha=0.05$, $\pi_0=0.2$ and $m_0 =2$, the upper bound in \eqref{eq30} becomes $0.02$,
and $n_{\ast}=147$, $148$ or $149$ validates \eqref{eq30} (whose corresponding left side quantity is $0.01928$, $0.01931$
or $0.01934$, respectively). Similar to \eqref{eq20}, we admit that condition \eqref{eq30} is restrictive.

\subsection{Tightening FDR bounds associated with mid p-values}

\label{secTighten}

In this section, we will derive potentially better FDR bounds for the BH
procedure applied to two-sided mid p-values. The discussion will use the
notations in \autoref{SecDefinition} and the beginning of
\autoref{SecTwoTypes}.

Let $X\sim\mathsf{Binomial}\left(  0.5,n\right)  $ with CDF $F$. Then $F$ is
symmetric with respect to $2^{-1}n$. On the other hand, for $X\sim
\mathsf{HGeom}\left(  1,\mathbf{N}\right)  $ with $\mathbf{N}=\left(
N,N,M\right)  $, its CDF $F$ is symmetric with respect to $2^{-1}M$. Let
$\check{x}$ be the smaller of the two modes of $f$ when $n$ or $M$ is odd, or
let $\check{x}$ be the mode of $f$ when $n$ or $M$ is even. Fix a $t\in\left(
0,1\right)  $. Then regardless of whether $X$ is $\mathsf{Binomial}\left(
0.5,n\right)  $ or $\mathsf{HGeom}\left(  1,\mathbf{N}\right)  $ with
$\mathbf{N}=\left(  N,N,M\right)  $,%
\[
\varpi\left(  x_{0}\right)  =l\left(  x_{0}\right)  +2^{-1}e\left(
x_{0}\right)  =2F\left(  x_{0}\right)  -f\left(  x_{0}\right)  \geq2F\left(
x_{0}-1\right)
\]
for $0\leq x_{0}<\check{x}$, and%
\[
\Pr\left(  \varpi\left(  X\right)  \leq t\right)  \leq\int_{\left\{  0\leq
x<\check{x}:2F\left(  x-1\right)  \leq t\right\}  }\mathsf{d}F\left(
x\right)  .
\]
Let $y\left(  t\right)  =\max\left\{  x\leq\check{x}:F\left(  x\right)  \leq
t\right\}  $. Then $y\left(  t\right)  <\check{x}$ and%
\[
\int_{\left\{  0\leq x<\check{x}:2F\left(  x-1\right)  \leq t\right\}
}\mathsf{d}F\left(  x\right)  \leq2^{-1}t+f\left(  y\left(  t\right)
+1;n\right)  \text{,}%
\]
i.e.,%
\begin{equation}
\Pr\left(  \varpi\left(  X\right)  \leq t\right)  \leq2^{-1}t+f\left(
y\left(  t\right)  +1;n\right)  . \label{eq55}%
\end{equation}

Employing the inequality (\ref{eq55}), we have the following:

\begin{theorem}
\label{ThmBetterBounds}Assume $\mathbf{N}_{i}=\left(  N,N,M_{i}\right)  $ and
the independence between $\left\{  \varpi_{i}\right\}  _{i=1}^{m}$. Then for
BT's and FET's, the FDR $\hat{\alpha}_{\mathsf{BH}}$ of the BH procedure
satisfies%
\begin{equation}
\hat{\alpha}_{\mathsf{BH}}=\sum_{i\in I_{0}}\sum_{r=1}^{m}\frac{1}{r}\left(
\tau_{r}+2^{-1}e_{i}\left(  x_{i}\left(  \tau_{r}\right)  \right)  \right)
\Pr\left(  C_{r}^{\left(  -i\right)  }\right)  \leq\hat{\alpha}_{1}%
+\hat{\alpha}_{2}, \label{eq52}%
\end{equation}
where $\sum_{r=1}^{m}\Pr\left(  C_{r}^{\left(  -i\right)  }\right)  =1$ for
any $i\in I_{0}$,%
\begin{equation}
\hat{\alpha}_{1}=2^{-1}\pi_{0}\alpha\text{ \ and \ }\hat{\alpha}_{2}%
=\sum_{i\in I_{0}}\sum_{r=1}^{m}\frac{f_{i}\left(  y_{i}\left(  \tau
_{r}\right)  +1\right)  }{r}\Pr\left(  C_{r}^{\left(  -i\right)  }\right)  .
\label{eq64}%
\end{equation}

\end{theorem}

The proof of \autoref{ThmBetterBounds} is straightforward from
\autoref{LmAntiCons} and omitted. The upper bound in (\ref{eq64}) may induce
less restrictive conditions than those required by \autoref{PropMTPBt} and
\autoref{PropMTPFET} in order to ensure the conservativeness of the BH
procedure when it is applied to two-sided mid p-values. In particular, the
FDR\ bound $\hat{\alpha}_{1}$ directly associated with the super-uniformity
part in the decomposition of the CDF of a two-sided mid p-value is reduced to
one half of $\pi_{0}\alpha$, and the remaining part $\hat{\alpha}_{2}$ can be
assessed by examining the behavior of each $f_{i}$ with respect to
$\boldsymbol{\tau}=\left(  \tau_{1},\ldots,\tau_{m}\right)  $. The strategy
presented above to obtain better FDR\ bounds can be generalized to multiple
testing where p-values have symmetric c\`{a}dl\`{a}g functions.

\section{Simulation study}

\label{SecSimuStudy}

In this section, we will numerically assess the performance of the BH
procedure and its adaptive version when they are applied to two-sided mid
p-values of BT's and FET's. Specifically, at a nominal FDR level $\alpha
\in\left(  0,1\right)  $, the adaptive BH\ procedure is implemented at nominal
FDR level $\alpha/\hat{\pi}_{0}$, where $\hat{\pi}_{0}$ is the estimator of
the proportion $\pi_{0}$ developed by \cite{Chen:2016c} that adapts to the
discreteness of p-values and reduces to the estimator in \cite{Storey:2004}
for continuous p-values. Note that this adaptive BH procedure has been shown
by \cite{Chen:2016c} to be conservative when it is applied to conventional p-values.

We will compare $\hat{\pi}_{0}^{\text{Convp}}$ and $\hat{\pi}_{0}%
^{\text{Midp}}$ obtained by applying $\hat{\pi}_{0}$ to mid p-values and
conventional p-value respectively, with $\hat{\pi}_{0}^{\text{Randp}}$, the
estimator obtained by applying Storey's estimator in \cite{Storey:2004} with
$\lambda=0.5$ to randomized p-values. We choose $\lambda=0.5$ for Storey's
estimator since other methods provided by the \textsf{qvalue} package to
implement this estimator severely under-estimates $\pi_{0}$ when it is applied
to randomized p-values. We will compare the procedure of \cite{Habiger:2015}
(denoted by \textquotedblleft SARP") that is obtained by applying Storey's
procedure in \cite{Storey:2004} with $\hat{\pi}_{0}^{\text{Randp}}$ to
randomized p-values, the adaptive BH procedure applied to conventional
p-values (\textquotedblleft aBH"), the adaptive BH procedure applied to mid
p-values (\textquotedblleft aBH-Midp), the BH procedure applied to
conventional p-values (\textquotedblleft BH\textquotedblright), and the BH
procedure applied to mid p-values (\textquotedblleft BH-Midp\textquotedblright).

\subsection{Simulation design}

The simulation, similar to that in \cite{Chen:2016c}, is set up as follows.
Set $m =20$, $10^3$ or $10^5$, $\pi_{0} = 0.5$, $0.6$, $0.7$, $0.8$ or $0.95$, $m_{0}%
=m\pi_{0}$, and nominal FDR level to be $0.05$. For each value for $\pi_{0}$,
do the following:

\begin{enumerate}
\item Generate Poisson and Binomial data:

\begin{enumerate}
\item Poisson data: let \textsf{Pareto}$\left(  l,\sigma\right)  $ denote the
Pareto distribution with location $l$ and shape $\sigma$ and $\mathsf{Unif}%
\left(  a,b\right)  $ be the uniform distribution on the interval $[a,b]$.
Generate $m$ $\theta_{i1}$'s independently from $\mathsf{Pareto}\left(
3,8\right)  $. Generate $m_{1}$ $\rho_{i}$'s independently from $\mathsf{Unif}%
\left(  1.5,6\right)  $. Set $\theta_{i2}=\theta_{i1}$ for $1\leq i\leq m_{0}$
but $\theta_{i2}=\rho_{i}\theta_{i1}$ for $m_{0}+1\leq i\leq m$. For each
$1\leq i\leq m$ and $g\in\left\{  1,2\right\}  $, independently generate a
count $\xi_{ig}$ from the Poisson distribution $\mathsf{Poisson}\left(
\theta_{ig}\right)  $ with mean $\theta_{ig}$.

\item Binomial data: generate $\theta_{i1}$ from $\mathsf{Unif}\left(
0.15,0.2\right)  $ for $i=1,\ldots,m_{0}$ and set $\theta_{i2}=\theta_{i1}$
for $i=1,\ldots,m_{0}$. Set $\theta_{i1}=0.2$ and $\theta_{i2}=0.6$ for
$i=m_{0}+1,\ldots,m$. Set $n=20$, and for each $g\in\{1,2\}$ and $i$,
independently generate a count $\xi_{ig}$ from
$\mathsf{Binomial}\left(  \theta_{ig},n\right)  $.
\end{enumerate}

\item With $\xi_{ig}$, $g=1,2$ for each $i$, conduct BT or FET to test
$H_{i0}:\theta_{i1}=\theta_{i2}\text{ versus }H_{i1}:\theta_{i1}\neq
\theta_{i2}$ and obtain the two-sided p-value $P_{i}$ of the test. Apply the
FDR procedures to the $m$ p-values $\left\{  P_{i}\right\}  _{i=1}^{m}$.

\item Repeat Steps 2. to 3. $250$ times to obtain statistics for the
performance of each estimator and FDR procedure.
\end{enumerate}

In addition to the independent data generated above, for $m=10^5$ positively and blockwise correlated Poisson and Binomial data are
generated as follows:

\begin{itemize}
\item Construct a block diagonal, correlation matrix $\mathbf{D}$ with $50$ equal-sized blocks, such that
for each block its off-diagonal entries are identically $0.1$. Generate a realization $\mathbf{z}%
=(z_{1},\ldots,z_{m})$ from the $m$-dimensional Normal distribution with zero
mean and correlation matrix $\mathbf{D}$, and obtain the vector $\mathbf{u}%
=(u_{1},\ldots,u_{m})$ such that $u_{i}=\Phi(z_{i})$, where $\Phi$ is the CDF
of the standard Normal random variable.

\item Maintain the same parameters used to generate independent Poisson and
Binomial data, and for each $g\in\left\{  1,2\right\}  $ and $i\in\left\{
1,\ldots,m\right\}  $, generate a count $\xi_{ig}$ corresponds to quantile
$u_{i}$ of the CDF of $\mathsf{Poisson}\left(  \theta_{ig}\right)  $ or
$\mathsf{Binomial}\left(  \theta_{ig},n\right)  $.
\end{itemize}

Note that the conditions of \autoref{PropMTPBt} and \autoref{PropMTPFET} are
not necessarily satisfied by the simulation design stated above.

\subsection{Summary of simulation results}

\label{secSimRes}

\begin{figure}[t]
\centering
\includegraphics[height=0.84\textheight,width=\textwidth]{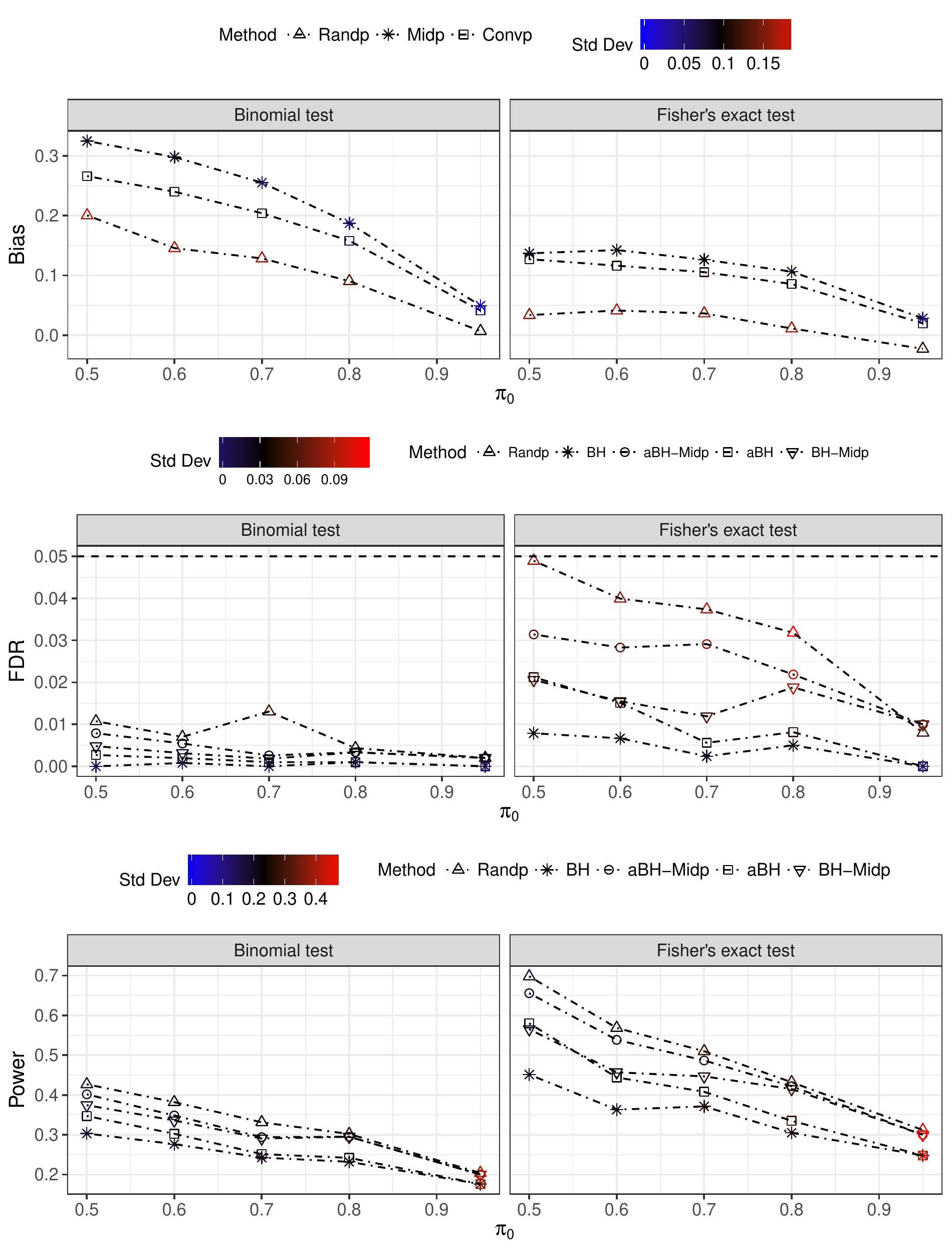}
\vspace{-0.5cm} \caption[sim independence]{Simulation results under independence when $m=20$. The top panel shows results of
estimating the true proportion $\pi_{0}$. ``Randp'' is $\hat{\pi}
_{0}^{\text{Randp}}$, ``Midp'' $\hat{\pi}_{0}^{\text{Midp}}$ and ``Convp''
$\hat{\pi}_{0}^{\text{Convp}}$. The legend ``Std Dev'' is the standard
deviation of each estimator. The middle panel shows the FDR of each procedure
where the legend ``Std Dev'' is the standard deviation of the FDP, and the bottom the power where the legend ``Std Dev'' is the
standard deviation of the TDP. 
}%
\label{figAlla}%
\end{figure}

\begin{figure}[t]
\centering
\includegraphics[height=0.84\textheight,width=\textwidth]{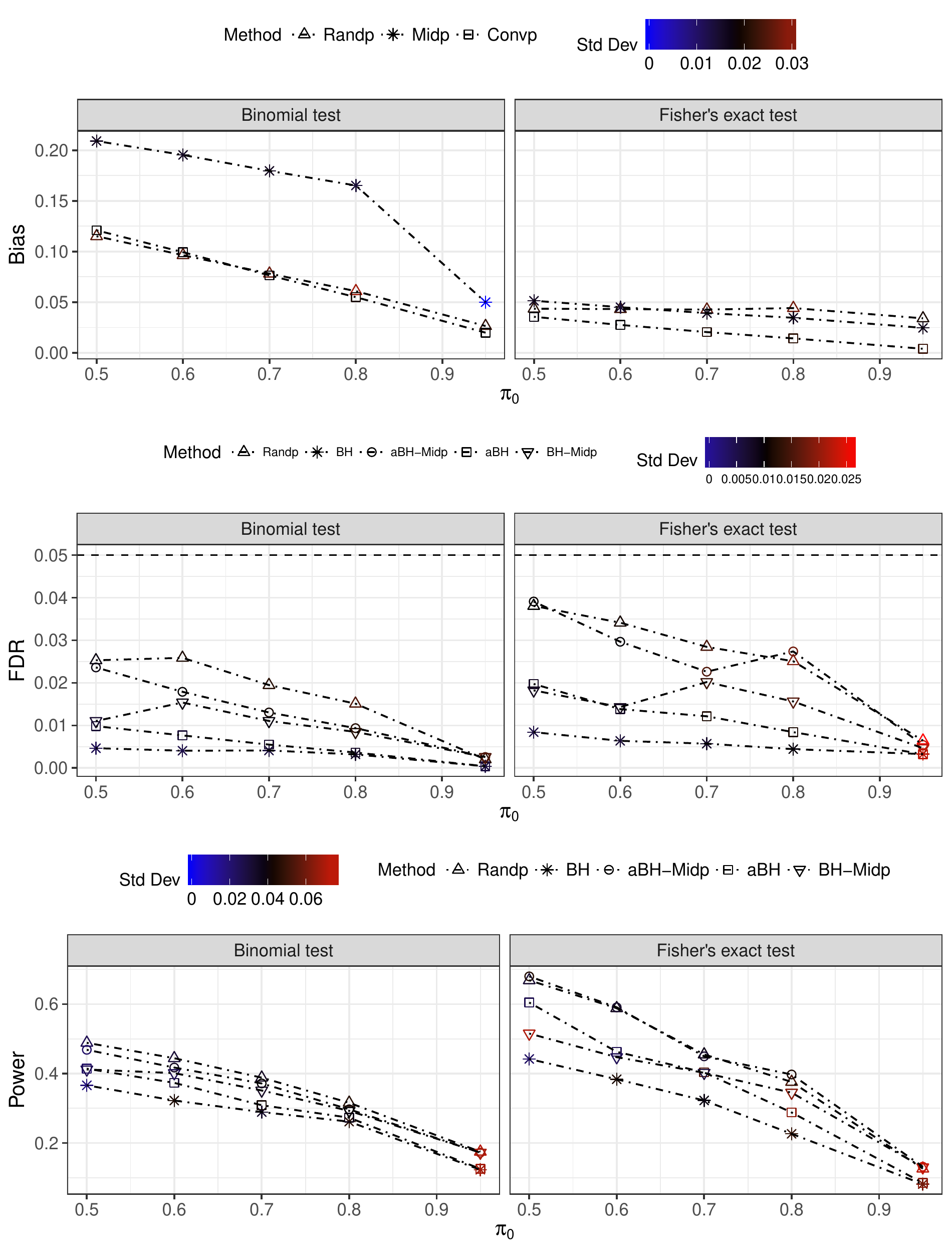}
\vspace{-0.5cm} \caption[sim independence]{Simulation results under independence when $m=10^3$. The top panel shows results of
estimating the true proportion $\pi_{0}$. ``Randp'' is $\hat{\pi}
_{0}^{\text{Randp}}$, ``Midp'' $\hat{\pi}_{0}^{\text{Midp}}$ and ``Convp''
$\hat{\pi}_{0}^{\text{Convp}}$. The legend ``Std Dev'' is the standard
deviation of each estimator. The middle panel shows the FDR of each procedure
where the legend ``Std Dev'' is the standard deviation of the FDP, and the bottom the power where the legend ``Std Dev'' is the
standard deviation of the TDP. 
}%
\label{figAll}%
\end{figure}

\begin{figure}[t]
\centering
\includegraphics[height=0.84\textheight,width=\textwidth]{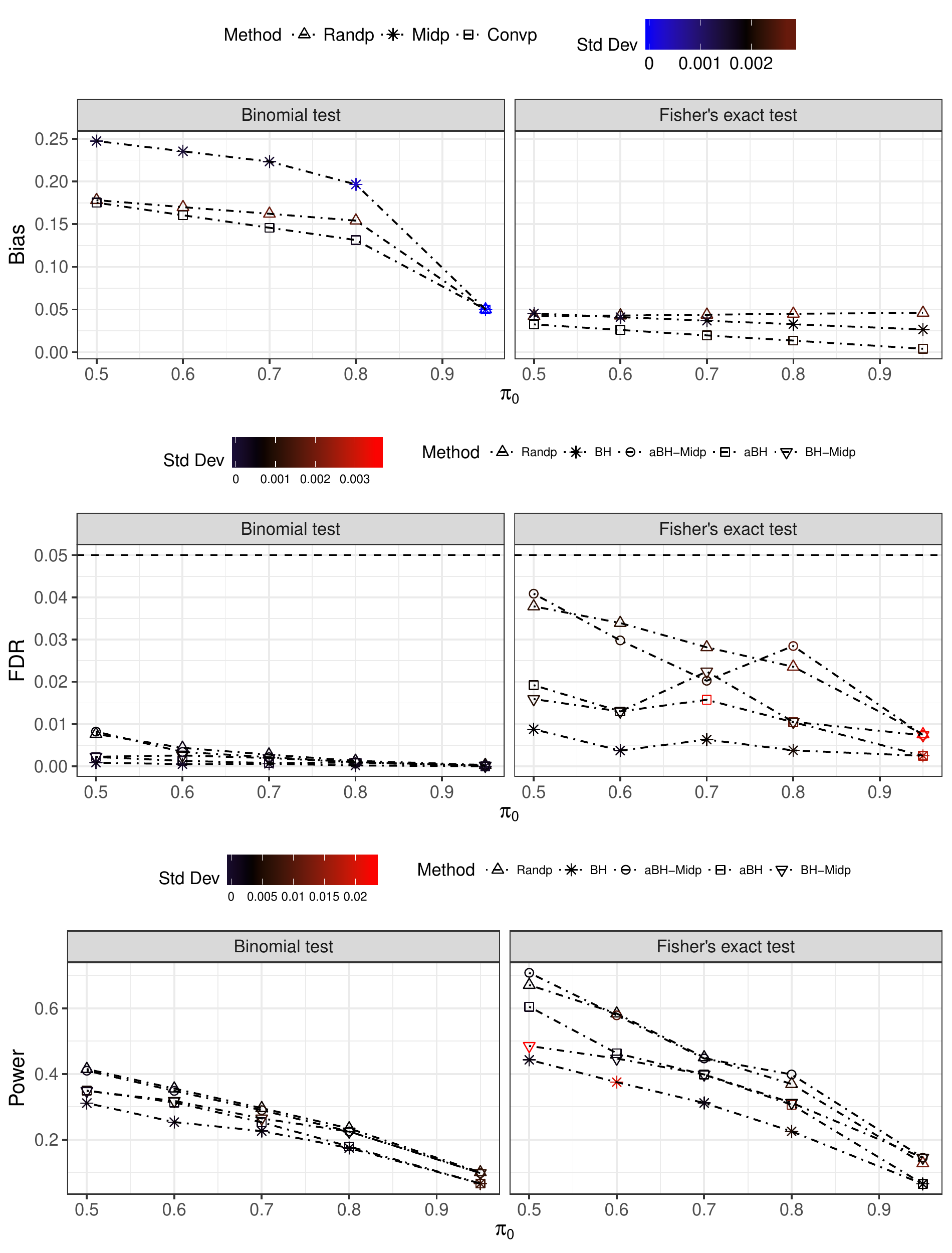}
\vspace{-0.5cm} \caption[sim independence]{Simulation results under independence when $m=10^5$. The top panel shows results of
estimating the true proportion $\pi_{0}$. ``Randp'' is $\hat{\pi}
_{0}^{\text{Randp}}$, ``Midp'' $\hat{\pi}_{0}^{\text{Midp}}$ and ``Convp''
$\hat{\pi}_{0}^{\text{Convp}}$. The legend ``Std Dev'' is the standard
deviation of each estimator. The middle panel shows the FDR of each procedure
where the legend ``Std Dev'' is the standard deviation of the FDP, and the bottom the power where the legend ``Std Dev'' is the
standard deviation of the TDP. 
}%
\label{figAllb}%
\end{figure}

\begin{figure}[t]
\centering
\includegraphics[height=0.84\textheight,width=\textwidth]{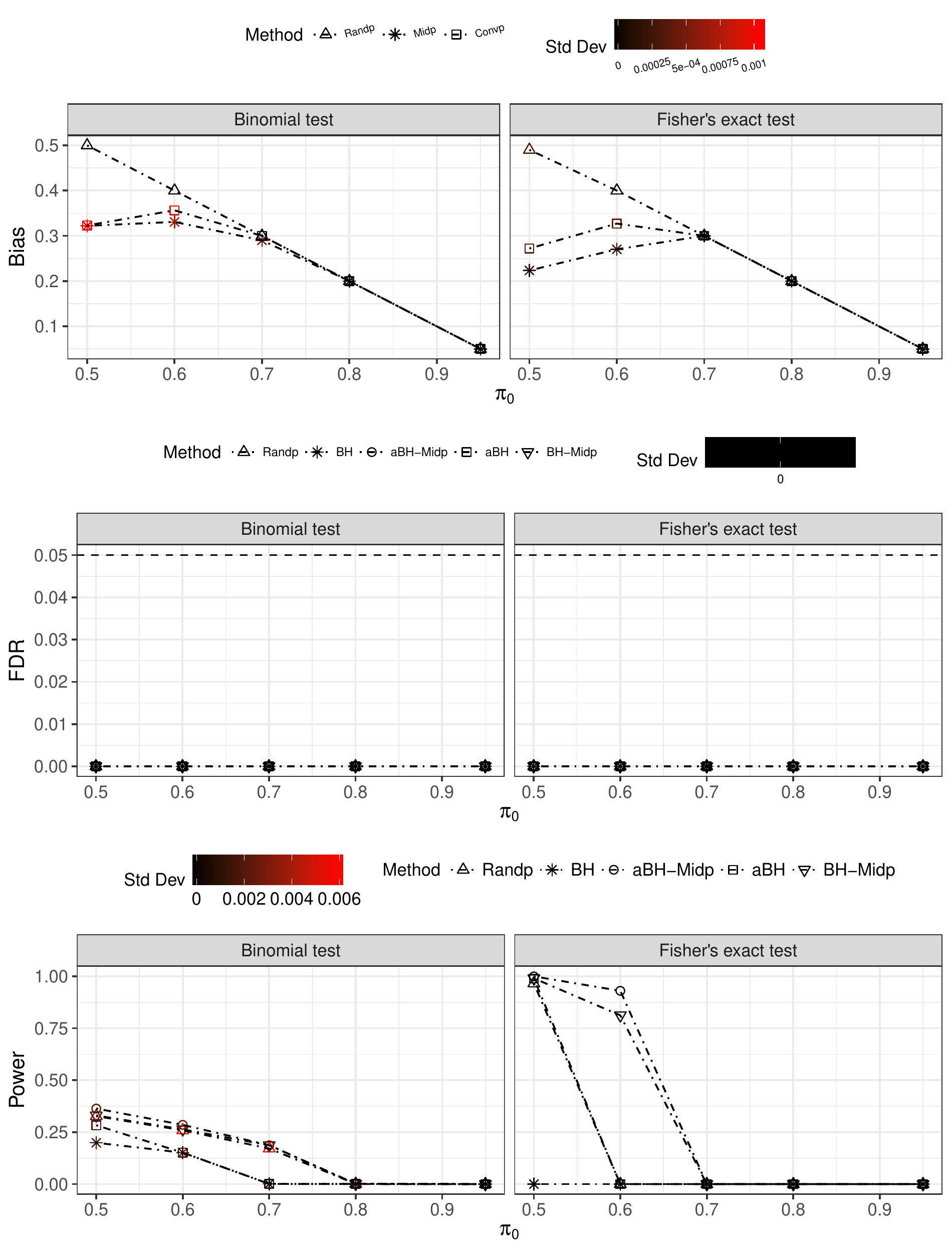}
\vspace{-0.5cm} \caption[sim independence]{Simulation results under positive, block dependence when $m=10^5$. The top panel shows results of
estimating the true proportion $\pi_{0}$. ``Randp'' is $\hat{\pi}
_{0}^{\text{Randp}}$, ``Midp'' $\hat{\pi}_{0}^{\text{Midp}}$ and ``Convp''
$\hat{\pi}_{0}^{\text{Convp}}$. The legend ``Std Dev'' is the standard
deviation of each estimator. The middle panel shows the FDR of each procedure
where the legend ``Std Dev'' is the standard deviation of the FDP, and the bottom the power where the legend ``Std Dev'' is the
standard deviation of the TDP. 
}%
\label{figAllc}%
\end{figure}

An estimator of the proportion $\pi_{0}$ is better if it is less conservative
(i.e., having smaller upward bias), is stable (i.e., having small standard
deviation), and induces a conservative adaptive FDR procedure. The top panels
of \autoref{figAlla}, \autoref{figAll}, \autoref{figAllb}, and \autoref{figAllc}
present the biases and standard deviations of the
estimators when they are applied to p-values of BT's or FET's. $\hat{\pi}_{0}$
applied to conventional p-values is stable and the most accurate among the
estimators, and $\hat{\pi}_{0}^{\text{SARP}}$ has relatively large standard
deviation. It is interesting to note that, for Binomial test, $\hat{\pi}_{0}$
applied to two-sided mid p-values may have relatively large bias when $\pi
_{0}$ is small.

We use the expectation of the true discovery proportion (TDP), defined as the
ratio of the number of rejected false null hypotheses to the total number of
false null hypotheses, to measure the power of an FDR procedure. Recall that
the FDR is the expectation of the false discovery proportion (FDP). We also
report the standard deviations of the FDP and TDP since smaller standard
deviations for these quantities mean that the corresponding procedure is more
stable in FDR and power. An FDR procedure is better if it is more powerful at
the same nominal FDR level and stable.

The middle and bottom panels of \autoref{figAlla}, \autoref{figAll}, \autoref{figAllb},
and \autoref{figAllc} record the FDRs and powers of
the procedures respectively. All procedures are conservative. Specifically, in the positive, blockwise dependence
setting in our simulation design, the FDRs of the procedures are very close to $0$,
whereas their powers can be close to $1$ when $\pi_0$ is considerably smaller than $1$ but are
very close to $0$ when $\pi_0$ is very close to $1$; see \autoref{figAllc}. This may be due to the clustering behavior
of signals or noise under positive, blockwise correlation for discrete data, and is worth further investigation.
The procedures
aBH-Midp and SARP have similar power performances and are the most
powerful among the procedures in comparison. aBH-Midp is stable but SARP seems
to be relatively less stable. The explanation for this is that the conditional
expectation of a randomized p-value is the corresponding mid p-value. So,
assuming that the $\hat{\pi}_{0}^{\text{SARP}}$ and $\hat{\pi}_{0}$ have
similar marginal distributions, the FDP and TDP of aBH-Midp and those of SARP
should have similar distributions after averaging out the extra uncertainty
induced by the uniform random variable in the definition of a randomized
p-value. Note that aBH and BH-Midp have similar power performances. An
explanation for this is that the improvement brought by $\hat{\pi}_{0}$ in the
adaptive BH procedure applied to conventional p-values can somehow be achieved
by applying the BH procedure to mid p-values since a mid p-value is smaller
than its corresponding conventional p-value.

\section{An application to HIV study}

\label{seAppAll}

We provide an application of the BH procedure based on two-sided mid p-values to multiple testing based
on discrete and heterogeneous p-value distributions in an HIV study. The naming conventions for the procedures compared in the simulation
study in \autoref{SecSimuStudy} will be used, and we will only compare BH, BH-Midp, aBH and aBH-Midp. All procedures are
implemented at nominal FDR level $0.05$

The study is well described in \cite{Gilbert:2005}.
The aim of the study is to identify, among $m=118$ positions, the
\textquotedblleft differentially polymorphic\textquotedblright\ positions,
i.e., positions where the probability of a non-consensus amino-acid differs
between two sequence sets. Two sequence sets were obtained from $n=73$
individuals infected with subtype C HIV (and are categorized into Group 1) and $n=73$
individuals with subtype B HIV (and are categorized into Group 2), respectively. How multiple testing is
set up based on two-sided p-values of FET's can be found in \cite{Gilbert:2005}, where each
position on the two sequence sets corresponds to a null hypothesis that
\textquotedblleft the probabilities of a non-consensus amino-acid at this
position are the same between the two sequence sets\textquotedblright.

There are $50$ positions for which the total observed counts are identically
$1$ and the corresponding two-sided p-value CDF's are Dirac masses. To reduce
the uncertainty induced by positions whose observed total counts are too small,
we only analyze those whose observed total counts are at least $2$. This gives
$68$ positions, i.e., $68$ null hypotheses to test. BH makes $15$ discoveries, BH-Midp $16$, aBH $16$ and aBH-Midp $25$,
showing the improvement that multiple testing based on mid p-values can bring. The additional
discoveries made by the procedures based on mid p-values are worth further investigation, had we been
able to prove their conservativeness.

\section{Discussion}

\label{SecDiscussion}

This paper is motivated by the scope of improving the BH procedure in
controlling FDR when it is applied to mid p-values, which has been realized by
researchers in multiple testing but no significant progress has been made yet
in investigating conditions under which such improvements can be achieved.
Considering this procedure with two-sided mid p-values in the contexts of
Binomial and Fisher's exact tests, we have been able to establish sufficient
conditions for its conservativeness and provide numerical evidence on its
superior performance under these conditions relative to its relevant
competitors. Even though these conditions are simple, they depend on the
unknown proportion of true null hypotheses. Our study reveals the critical
role of this proportion in FDR control for a step-up procedure when p-values
are not super-uniform. The conservativeness of the BH procedure based on two-sided mid p-values
is also partially due to the existence of sub-intervals on which such a p-value is strictly super-uniform.

Since in practice we often have some information on at least how large the
proportion of true nulls is, based on inequality (\ref{eq67}), we can rescale
the critical constants of the BH procedure so that the modified procedure controls FDR. However, such
rescaling very likely will make the critical constants overall smaller than
$\left\{  i\alpha/m\right\}  _{i=1}^{m}$, thus potentially counterbalancing
the gain in power of applying the modified BH procedure to mid p-values. In other words, for the
multiple testing scenarios considered in this work, it is quite feasible to directly modify
the BH procedure to maintain FDR control for mid p-values but possibly at the
expense of unimproved power. On the
other hand, to develop more powerful MTP's based on mid p-values whose
conservativeness is ensured under weaker conditions than we have presented, a
tighter estimate of
\begin{equation}
\xi_{i}=\sum_{r=1}^{m}\frac{e_{i}\left(  x_{i}\left(  \tau_{r}\right)
\right)  }{r}\Pr\left(  C_{r}^{\left(  -i\right)  }\right)  ,i\in I_{0},
\label{eq19}%
\end{equation}
than given in this paper is needed but usually very hard to obtain. We leave
this to future research.

\bibliographystyle{elsarticle-num}

\clearpage

\end{document}